\documentclass[12pt, twoside, here]{article}
\usepackage{epsf}
\usepackage{times,colordvi,amsmath,epsfig,float,color,multicol}
\usepackage{graphics}
\usepackage{hhline}
\usepackage[large]{subfigure}
\usepackage[latin1]{inputenc}
\usepackage{listings}

\title{A Rotatable Stabiliser for the Control of Pitch and Yaw in a Radio-controlled Aircraft}

\author{S. J. Childs, M. H. Jackson\footnotemark[2] \ and I. Basson\footnotemark[2] \\ \\ {\small\em Department of Mathematics and Applied Mathematics} \\ {\small\em \footnotemark[2] Division of Electronics} \\ {\small\em University of the Free State, P.O. Box 339, Bloemfontein, 9300, South Africa.} \\ {\small\em tel: +27 728459556, email: simonjohnchilds@gmail.com}}

\date{}       

\begin{document}

\maketitle

\begin{abstract}
\noindent {\em This research implements a design for the control of a rotatable
stabiliser which, it is proposed, might augment, or fully replace, the
conventional control mechanisms for pitch and yaw in certain types of aircraft.
The anticipated advantages of such a device are around 25\% less drag, for a
capability which ranges between equivalent and greater than twofold that of a
conventional tail. The tail of a popular, radio-controlled, model aircraft is
replaced with a rotatable stabiliser and its rotation is effected by way of a
continuously-rotating servo, modified with a potentiometer. Two hollow,
carbon-fibre shafts (one sleeved within the other so as to allow free rotation)
serve as the mechanical link between the servo and the tail. Inserting wiring
along the full length of the innermost shaft and incorporating three slip rings
into its collar, at the forward, servo end, facilitated an electrical supply to
the tail. A mapping betweeen the position of the controls and states of the
stabiliser is formulated. Small and continuous adjustments cause the stabiliser
to rotate in the opposite direction to the controls (when viewed from aft) and
the deflection of the hinged control surface is proportional to the radial
displacement of the controls from their centred position. For what would amount
to large and contradictory rotations of the device in terms of this protocol
(rotations greater than 90~$^\circ$), the desired configuration can be more
efficiently achieved by regarding the device's original orientation to differ by
180~$^\circ$ from what it actually is and by reversing the sign of the
deflection. One consequence of this latter mode of control is that a symmetrical
aerofoil is indicated. General flight, including static longitudinal and
directional stability, was not found to be compromised.}
\end{abstract}


\section{Introduction}

This research sets out to implement a design for the control of a rotatable
stabiliser, an aerofoil whose in-flight orientation is actively adjusted by
swivelling around an axis parallel, or approximately parallel, to the aircraft's
longitudinal axis (Childs \cite{patentChilds11}, \cite{patentChilds12} and
\cite{patentChilds13}). The envisaged purpose of the device, is to augment, or
fully replace, the conventional control mechanisms of pitch and yaw in certain
types of aircraft. The anticipated advantages of a rotatable stabiliser are
around 25\% less drag, for a capability which ranges between equivalent and
greater than twofold that of the conventional tail. Interference drag should
also be eliminated. One, anticipated handicap of the device is the potential for
it to stall, from its tips, inward, if rotated too fast. Civil aeroplane designs
must, furthermore, demonstrate inherent static longitudinal and directional
stability. 

\begin{figure}[H]
    \begin{center}
        \includegraphics[width=14cm, angle=0, clip = true]{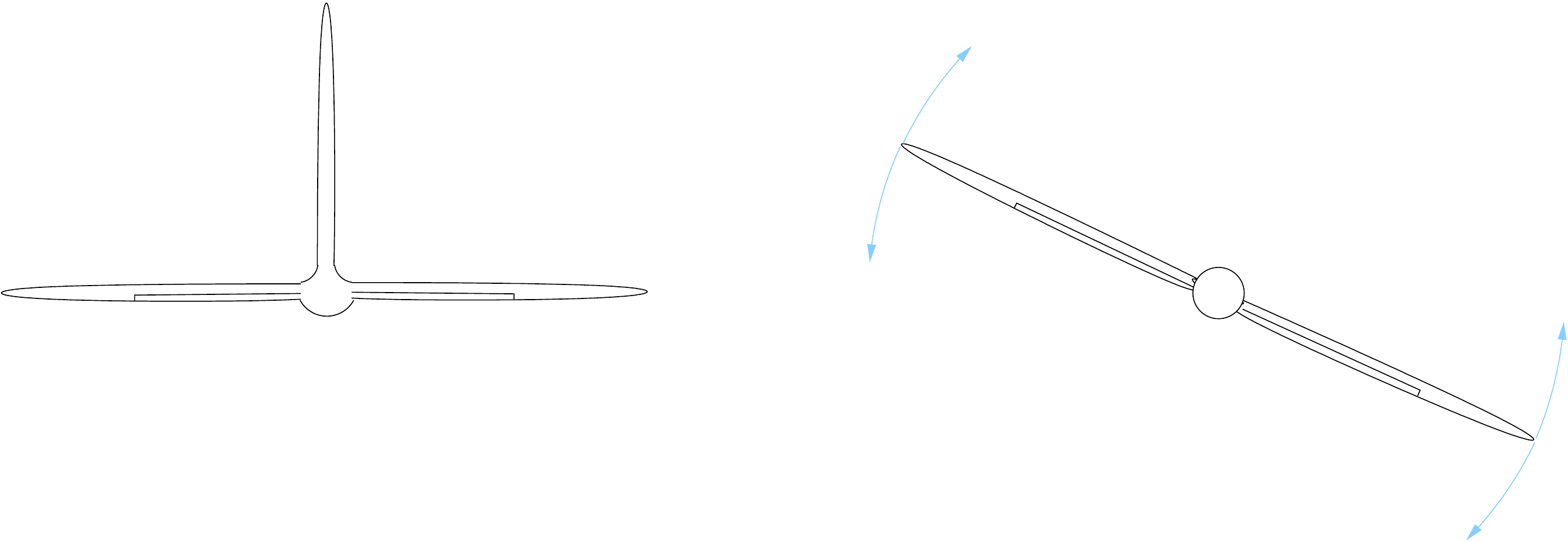}
	\caption{A conventional tail and a rotatable stabiliser, as viewed down
the longitudinal axis.} \label{profile}
   \end{center}
\end{figure}

Conventional aircraft tails consist of a horizontal and vertical stabiliser at
right angles to each other, as well as their respective control surfaces, the
elevator and rudder. The elevator provides for the control of pitch, while the
rudder provides for the control of yaw. Both the pitching and yawing forces
combine to give a resultant force, the magnitude of which is given by
Pythagoras' theorem. The magnitude of the resultant is always less than the sum
of its pitching and yawing components, a fact which infers that the same force
can be achieved with a lesser area of aerofoil. It is for this reason that a
more efficient device is sought. A single aerofoil in the correct orientation
will always be able to produce the same lift force while incurring a much lower
penalty in drag. It is for this reason that deploying a rotatable aerofoil (Fig.
\ref{profile}), precisely in the direction in which lift is required, is
contemplated as an alternative to the conventional tail. 

The function of the vertical stabiliser, is to maintain the aircraft's static
directional stability, while rudder inputs cause the aircraft to yaw. Yawing an
aircraft induces a secondary, rolling moment. It causes the outside wing to
speed up, to meet the remote, free air flow at a higher incident angle
(dependent on sweepback) and, consequently, to lift. It also causes the inside
wing to slow down, to meet the remote, free air flow at a lower incident angle
(again, dependent on sweepback) and be slightly shielded by the fueselage,
thereby losing lift. The aircraft ultimately rolls in the direction of yaw,
which happens to be the same direction as the bank the yaw is intended to
combine with in a coordinated turn. Of course, when the tail weathervanes, the
oscillatory yaw gives rise to an oscillatory version of this yaw-to-roll
phenomenon, known as Dutch roll. Dutch roll is particularly problematic in
wide-bodied, multi-engined aircraft with pronounced sweepback and a poorly
defined, variable centre of gravity; hence higher and offset moments of inertia.
Dutch roll is, nonetheless, often considered a small price to pay for the highly
desireable stall characteristics produced when sweepback is combined with
washout towards the wingtips. It is for this reason that a pronounced vertical
stabiliser is often a feature of such aircraft and active yaw damping, by way of
the rudder, may even be required. For smaller, unmanned, single-engined aircraft
with a well defined and constant centre of gravity and little or no sweepback,
the vertical stabiliser is of diminished importance. Examples of such aircraft
would include drones, U.A.Vs. and missiles. Wingtip modifications designed to
counter induced drag might be predicted to diminish the vertical stabiliser's
significance in future aircraft designs.

The rudder, itself, ordinarily gives rise to an adverse rolling moment about the
longitudinal axis. This is due to its asymmetrical positioning relative to that
axis, its superior position in the conventional tail configuration. The adverse
rolling moment never manifests itself, however, since it is over-compensated for
by the aforementioned, yaw-induced rolling moment. Ordinary v-tails, such as the
V35 Beechcraft Bonanza, suffer from adverse roll too. Only inverted v-tails
(such as that of the Predator drone) and tails in which the vertical stabiliser
is positioned below the longitudinal axis do not suffer from adverse roll. They
give rise to a moment which reinforces both the yaw-induced roll, as well as the
bank the yaw is intended to combine with in a coordinated turn. Such tails have,
what might therefore be termed, an advantageous roll as a bi-product. Since the
tail mostly exerts a downward force during flight, an anhedral or inverted
v-tail has, furthermore, the same levelling tendency as a dihedral wing, whereas
the ordinary v-tail has a destabilising effect. A rotatable stabiliser should
produce no adverse roll once in the correct orientation. This prediction is
based on its proposed symmetrical arrangement about the longitudinal axis. A
moment of adverse roll may, however, arise during the rotation of the device;
unless seperate control surfaces (or the angles of attack) on either side of the
axis, are adjusted in the direction of rotation. 

The primary function of the horizontal stabiliser and its control surface, the
elevator, is to counter the gravitational moment about the centre of pressure
associated with the wing. A centre of gravity forward of the centre of pressure
is an arrangement essential to the fundamental stall characteristics any
aircraft must possess. Since the elevator is used to adjust the pitch of the
aircraft and, therefore, the angle of attack of its wings, a secondary function
of the elevator is to determine the speed of the aircraft as well as the lift
force, consequently the rate of climb, or descent. When the airspeed drops below
that required for the wing to produce sufficient lift, the instinctive
pilot-response is for the elevator to be adjusted so as to cause the angle of
attack of the wings to exceed the stalling angle. Under such circumstances, the
wing is no longer a streamlined body and, since Bernoulli's equation only
applies along a streamline, the wings stall. In comparison, the rudder is mostly
used to add yaw while simultaneously banking. This combination of bank and yaw
produce either a coordinated turn or, should the controls be crossed, a
side-slip. An insufficient rudder input will cause the aircraft to slip in the
turn, while an excessive rudder input will cause it to skid. Yawing an aircraft
at low airspeeds can result in the inside wing stalling and the aircraft then
enters a spin, the remedy for which is an opposite rudder input. In summary, the
horizontal stabiliser may be considered more important than the vertical
stabiliser, barring certain aerobatic applications and spin recovery. One
variation on the horizontal stabiliser theme is worthy of mention, namely the
stabilator. The stabilator is a horizontal stabiliser in which the entire
stabiliser becomes the control surface. An adjustable angle of attack allows the
entire horizontal stabiliser to perform the function of the elevator.

What is ultimately envisaged in this research is an aerofoil whose in-flight
orientation is actively adjusted by
swivelling\footnotemark[1]\footnotetext[1]{Note the capability implied by the
use of the operative word ``swivelling'' as opposed to ``tilting''.} relative to
the aircraft, around an axis parallel, or approximately parallel, to the
longitudinal axis (an axis designed to be approximately parallel to the relative
air flow), thereby affording the aerofoil the capability of exerting a lift
force in any direction about the said axis. The direction of this lift force
can, furthermore, be affected continuously, for all changes. It is anticipated
that such a device may be used either aft or forward of the wing and centre of
gravity. That is, in either a tail or canard-wing position. The purpose of the
envisaged device, is to augment, or fully replace, the conventional control
mechanisms of pitch and yaw in aircraft, although more ambitious applications
are not excluded. For example, the possibility of implementations in which the
device  might also augment, or fully replace, the function of the ailerons (a
l\'{a} elevons) cannot be ignored, should the device have seperate, hinged
control surfaces, on either side of the axis about which it rotates (or a split
device in the instance of a stabilator-like implementation). Variations on this
theme therefore may, or may not, include a split, variable pitch, a mechanism
for warping, or similar, for rapid rotation at low airspeeds.

In order to explore the concept further, the conventional tail of a popular,
radio-controlled, model aircraft was replaced with a rotatable stabiliser, the
electro-mechanical controls of which had to be designed and implemented. To
control the orientation of the stabiliser, a continuously-rotating servo was
modified by way of a position encoder (variable resistor). This then fullfilled
the function of a light-weight stepper motor, since one of sufficiently low
mass, size and high torque could not be sourced. It was mounted as close to the
aircraft's original centre of gravity as possible, so as to preserve the
aircraft's fundamental stall characteristics. The tail and servo were connected
mechanically by way of a hollow, carbon-fibre shaft, contained within a second,
carbon-fibre sleeve. The sleeve allowed the hollow shaft to rotate freely and
smoothly inside the tail-boom. Wiring was inserted and run along the inside of
the hollow shaft, for its full length. Three slip rings were added at its
forward, servo-end and three, corresponding, spring-loaded contacts were mounted
on an arm attached to the servo, thereby completing an electrical supply to the
tail (a second servo on the tail is needed to control the deflection of the
stabiliser's control surface). For succinctness, the function of the joystick
traditionally assigned to the control of ailerons was initially replaced by that
traditionally associated with the rudder pedals. Its pitch-controlling function
was otherwise designed to be conventional. Convention dictates that, in the
absence of any lateral displacement of the joystick, the orientation of the
stabiliser and the deflection of its control surface should correspond to those
of a horizontal stabiliser and its elevator, respectively. This means that for
purely lateral displacements of the joystick, the orientation of the stabiliser
and the deflection of its control surface correspond to those of a vertical
stabiliser and rudder, respectively. From these embedded substructures it
follows that small and continuous adjustments should cause the stabiliser to
rotate in the opposite direction to the joystick (when viewed from aft) and the
deflection of the hinged control surface is proportional to the radial
displacement of the joystick from its centred position. For what, in terms of
that protocol, would amount to large and contradictory rotations of the device
(rotations greater than 90~$^\circ$), the desired configuration can be more
efficiently achieved by regarding its original orientation to differ by
180~$^\circ$ from what it actually is and by reversing the sign of deflection.
Once a continuous, one-to-one, conformal mapping, between the position of the
controls and the rotatable stabiliser is devised, the more traditional
assignment of the controls can be reverted to for the practical purposes of
flight testing.

\section{The Advantage Over a Conventional Tail}

\begin{figure}[H]
    \begin{center}
        \includegraphics[width=14cm, angle=0, clip = true]{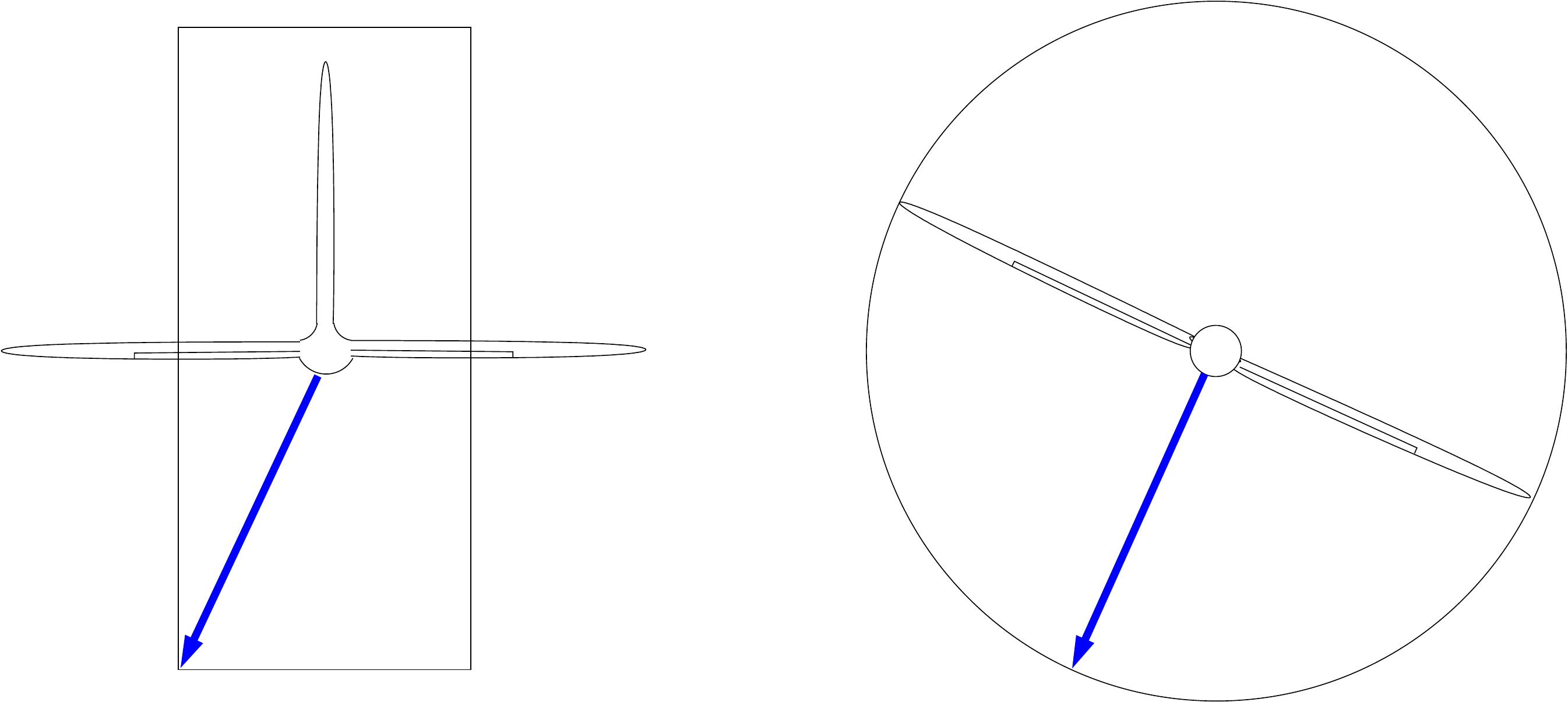}
 	\caption{Envelopes of maximum force for the conventional tail and rotatable stabiliser respectively.} \label{envelopeMaxForce}
   \end{center}
\end{figure}

In the case of a conventional aircraft tail the envelope of maximum force is
rectangular about the aircraft's longitudinal axis (Fig.
\ref{envelopeMaxForce}), it being limited to exert its maximum in only four,
unique directions. In contrast, the rotatable stabiliser is able to exert this
maximum in any direction, the envelope of maximum force being circular (Fig.
\ref{envelopeMaxForce}). 
Consider the capability of a convevtional tail, in which the vertical and
horizontal stabiliser are, for simplicity, comprised of three identical members, as depicted in Fig. \ref{envelopeMaxForce}. If the area of each member is
$S$, then the maximum resultant is obtained by combining the maximum force
exerted by both the horizontal and vertical stabilisers simultaneously, that is
\begin{eqnarray} \label{equivalentArea}
\sqrt[2]{\left( \frac{1}{2} \rho \ C_L v^2 \ S \right)^2 + \left( \frac{1}{2} \rho \ C_L v^2 \ 2 S \right)^2} &=& \frac{1}{2} \rho \ C_L v^2 \ S \ \sqrt[2]{1 + 2^2} \nonumber \\
&=& \frac{1}{2} \rho \ C_L v^2 \ 2.236 \ S, 
\end{eqnarray}
in which $\rho$, $C_L$ and $v$ are the usual, density of the air, coeficient of
lift and velocity of the incident air, respectively. 

The above result (Eq. \ref{equivalentArea}), is immediately recognizeable as
the lift formula for an aerofoil of area $2.236 \times S$. In this way it
becomes clear that only $2.236 \times S$ of correctly-orientated aerofoil is
required to produce the same maximum lift as a $3 \times S$ area of conventional
tail. A far smaller area of aerofoil than the combined area of the horizontal
and vertical stabiliser is required to produce the same force. A reduction in
drag of approximately 25\% is therefore one consequence of resortng to a
rotatable stabiliser. The maximum capability is, furthermore, unrestricted in
the case of the rotatable stabiliser, whereas the conventional tail is only able
to attain this maximum in four, unique directions. The only four directions for
which the conventional tail under consideration fairs this well are 
\begin{eqnarray*}
\pm \left[ {\mathop{\rm arctan}}\left( 2 \right) + n \pi \right] &=& \pm \left( 63\,{}^\circ \times \frac{\pi}{180\,{}^\circ } + n \pi \right) \hspace{10mm} n = 0, 1, \cdots
\end{eqnarray*}
radians. For a purely yaw-related requirement, the rotatable device is able to
exert more than twice the force of the conventional tail, by Eq. \ref{equivalentArea}. A rotatable stabiliser therefore has the advantage of
being able to exert the maximum lift force in any direction, which can amount to
more than twice the capability, for a much reduced drag. Of course, the
individual members of a conventional tail do not operate independently from an
aerodynamic point of view, they operate rather as a single system, the
respective flows over each surface interacting with each other. The proposed
device might therefore also be expected to facilitate a lower interference drag,
there being two less intersecting surfaces. 

Such an analysis is, of course, a gross over simplification and the significance
of tail drag, itself, needs to be put into perspective. Drag ordinarily depends
on the flight regime, the percentage of laminar flow, etc. and induced drag can
also become a factor, depending on the speed of the aircraft. The relatively low
aspect ratio of the device, preferred for rotation, is a disadvantage from an
induced drag point of view. When it comes to parasitic drag, however, a close in
engine installation, gear doors and a plethora of other factors are by far the
greatest budget of drag on the airframe, the largest contributors to the overall
drag. Interference drag, cooling drag and propeller effects in the absence of
laminar flow are just a few of the other issues which bring the significance of
tail drag into perspective. 

Possible disadvantages of a rotatable device are that there is a limit on
response time, as will be shown, and a rotating link might exceed the mass of
the conventional system of cables and pulleys. In a manned aircraft the response
of the normal tail configuration is instantaneous, at least in so far as the
human input is. In a remotely controlled aircraft one relies on servos etc.
anyway.

\section{Mechanical and Electrical Modification of a Radio-controlled, Model Aircraft}

\begin{figure}[H]
    \begin{center}
        \includegraphics[width=15cm, angle=0, clip = true]{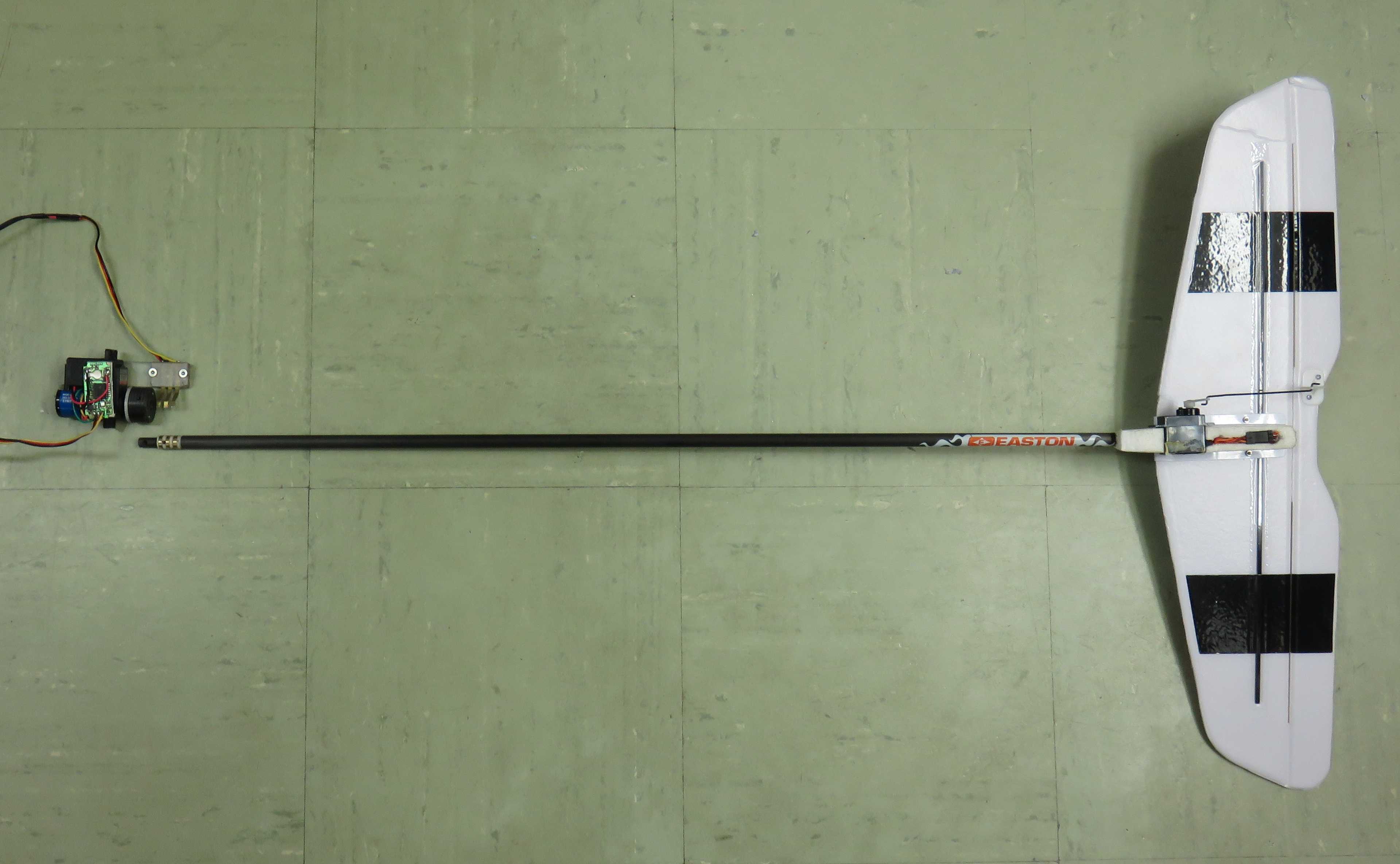}
	\caption{The machanics of the tail system: The mainly-visible, carbon-fibre sleeve contains a second, freely-rotating, carbon-fibre shaft within it (and in whose hollow centre the three wires, to supply the tail servo, were placed).}
\label{tailShaft}
   \end{center}
\end{figure}

The vertical stabiliser of a popular, radio controlled, model aircraft (the
Radian Pro) was removed and its tail-boom drilled out. A hollow, carbon-fibre
sleeve, which was chosen to accomodate a second, inner, carbon-fibre shaft in
such a manner as to allow its free rotation, was inserted into the tail boom.
The original horizontal stabiliser was attached to the back of the internal,
rotating shaft while the front of the shaft was attached to a modified servo at
the forward, cockpit end (refer to Fig. \ref{tailShaft}). The carbon-fibre
shafts allowed the modified servo to be positioned as close to the aircraft's
original centre of gravity as possible (so as to preserve the aircraft's
fundamental stall characteristics) while simultaneously providing a mechanical
connection to the tail.

\begin{figure}[H]
    \begin{center}
        \includegraphics[width=15cm, angle=0, clip = true]{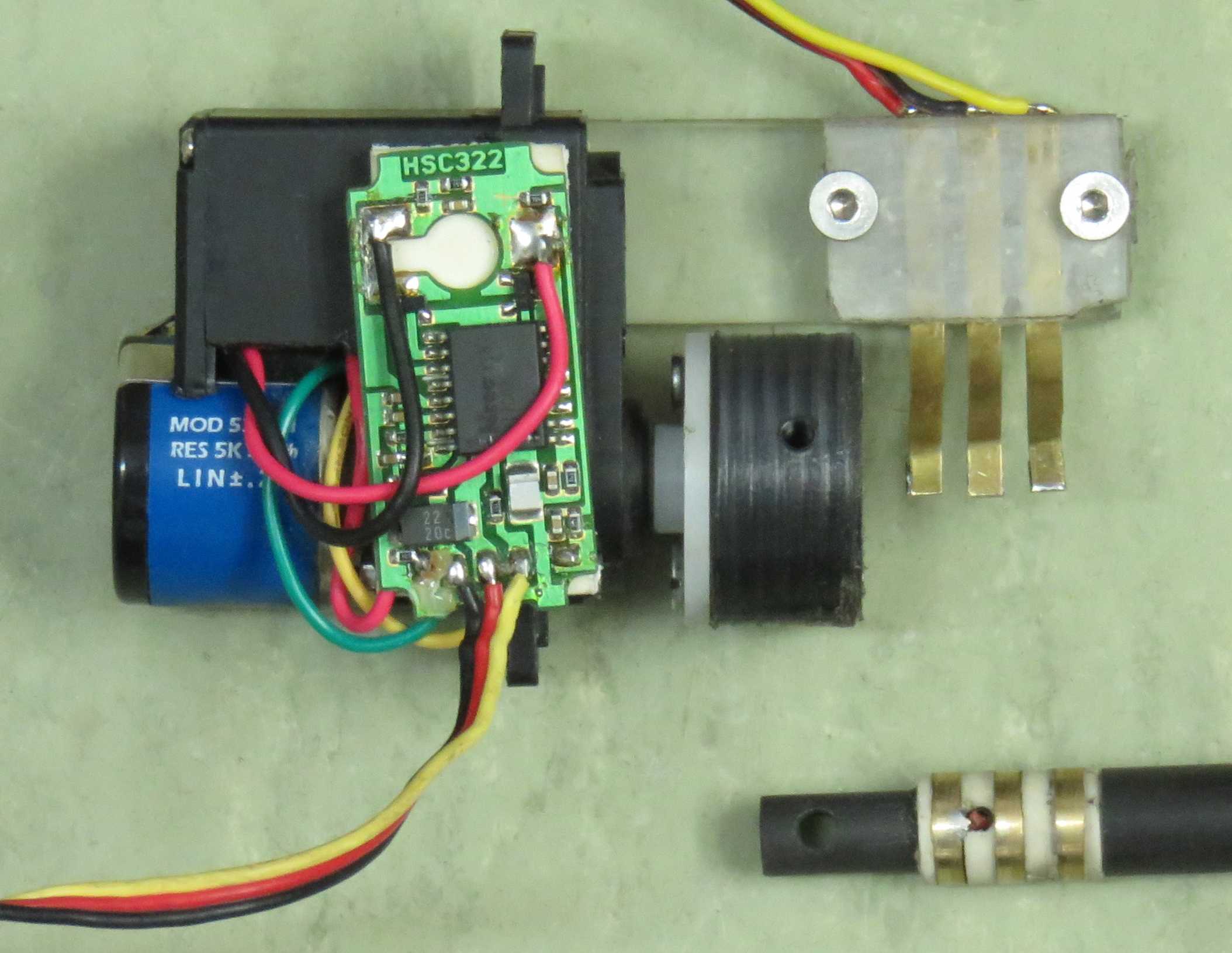}
	\caption{A continuously-rotating servo, modified to incorporate a
5~$\mathrm{k}\Omega$, three-turn, multi-turn potentiometer. Other modifications
include a chuck and an arm with three, sprung electrical contacts. The tip of
the carbon-fibre shaft (lower right) is inserted into the chuck and held in
place with a grub screw, thereby facillitating electrical contact between the
three slip rings and the spring-loaded contacts (the wiring for the tail is led
along the hollow centre of the shaft). Note that the white collar, featuring the
three slip rings, is followed immediately by a second carbon-fibre sleeve,
enveloping the shaft.} \label{modifiedServo}
   \end{center}
\end{figure}

A stepper motor of sufficiently low mass, size and high torque could not be
sourced and so a continuously-rotating servo, modified by way of a position
encoder, or variable resistor, was used as a light-weight substitute to fulfill
the task of rotating the stabiliser (refer to Fig. \ref{modifiedServo}). The
effect of incorporating a 5~$\mathrm{k}\Omega$, three turn, rotary potentiometer
produced a 180~$^\circ$ clockwise and 180~$^\circ$ counter-clockwise rotation,
with feedback. The rotating head of the modified servo was altered to feature a
chuck, designed to accomodate the tip of the carbon-fibre shaft, held in place
with a grub screw. Three, spring-loaded electrical contacts were also mounted on
an arm attached to the servo.

Inserting wiring along the full length of the innermost, hollow, carbon-fibre
shaft and incorporating three slip rings into its collar, so as to push against
the three, spring-loaded electrical contacts, facilitated an electrical supply
to the tail (refer to Fig. \ref{modifiedServo}). The purpose of the electrical
wiring is to supply the servo which controls the deflection of the stabiliser's
hinged, control surface. 
\begin{figure}[H]
    \begin{center}
        \includegraphics[width=15cm, angle=0, clip = true]{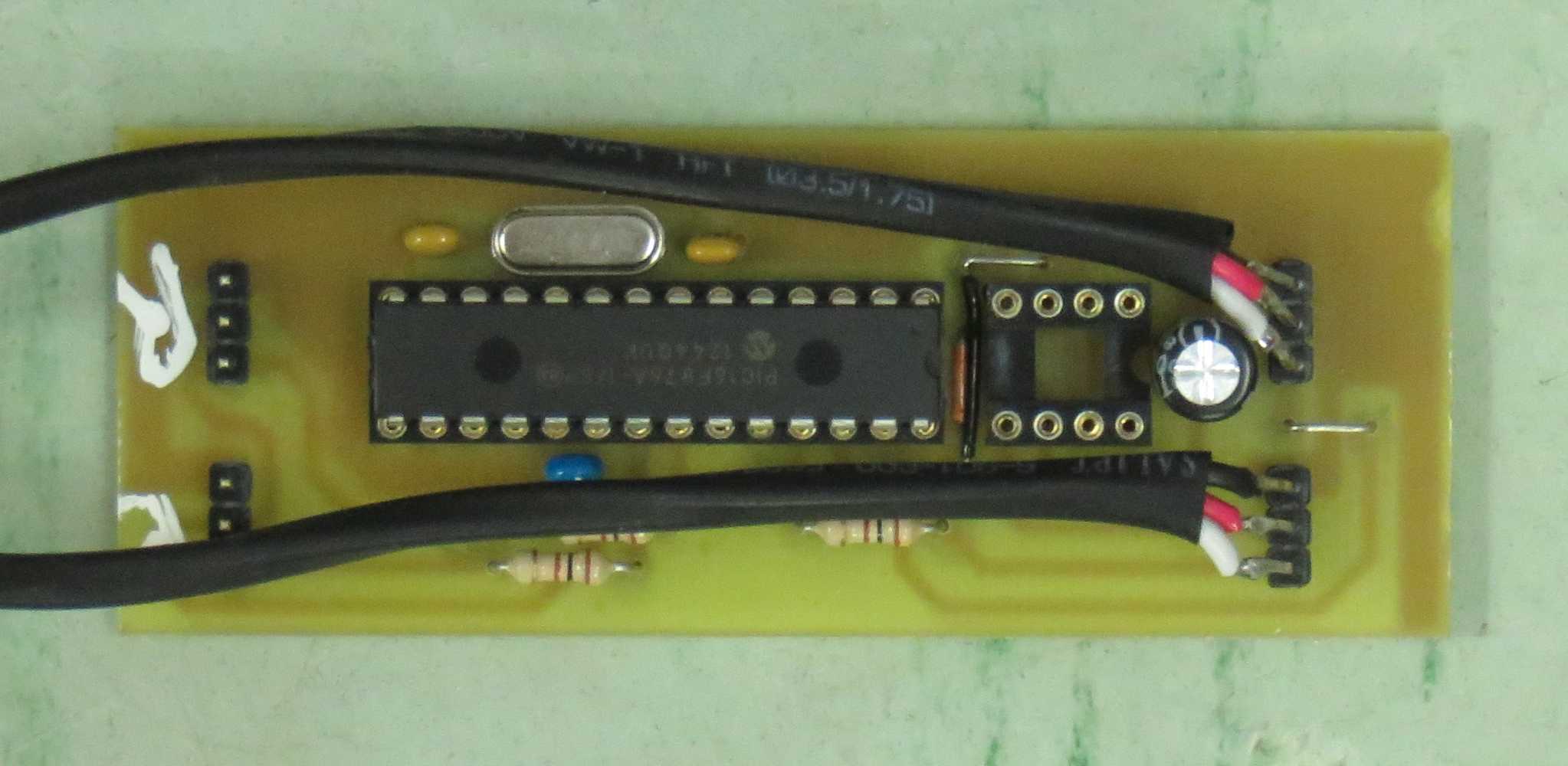}
	\caption{The circuit to which the RC signal is passed for processing, from the RC receiver. The 28-pin, dual inline package (DIP), visible in the centre of the circuit board, is the Microchip PIC16F876 microcontroller. It converts a position of the joystick into an orientation and deflection of the rotatable stabiliser. The vacant, eight-pin IC socket allows the connection of a debugger.}
\label{integratedCircuit}
   \end{center}
\end{figure}    

A standard, remote-control (RC) radio receiver was used to receive the signal
from the transmitter. It passes the transmitter-joysticks' position, by way of a
pulsed width modulation (PWM) signal, to the integrated circuit (IC). The
integrated circuit has multiple IO lines, 14~$\mathrm{kB}$ of programme memory
and a Microchip PIC16F876 microcontroller (Fig. \ref{integratedCircuit})
computes the desired output signals for the orientation of the rotatable
stabiliser and the deflection of its control surface. An eight-pin IC-socket was
included in the circuit to facilitate the connection of a debugger. The circuit
has two inputs and two outputs. Both inputs are connected to the radio receiver.
The microcontroller converts information on the joystick's position into an
orientation of the stabiliser and deflection of its control surface. This is, in
turn, converted to two outputted signals, one for each servo. The outputted
signals are sent to the respective servos which control the orientation and
deflection. The output signals use the same PWM principle as the input signals.
They consist of a high pulse followed by a low pulse, the total duration of
which is 20~$\mathrm{ms}$, from start to start. The duration of the high pulse
varies from a minimum of 1~$\mathrm{ms}$, to a maximum of 2~$\mathrm{ms}$.    

It would have been preferable, though not practical, to place the stabiliser
higher up, further out of the downwash. A low and far back horizontal stabiliser
has been implicated as a cause of flat spins by some Langley, spin-tunnel tests.
      
A summary list of the equipment, mechanisms and materials used in the modification would therefore be as follows:
\begin{enumerate}
\item Radian Pro Glider.
\item Continuously rotating servo
\item Vishay wirebound potentiometer (533 Series, three turn, 5~$\mathrm{k}\Omega$).
\item Microchip PIC16F876 microcontroller.
\item Two carbon-fibre arrow shafts (one to fit within the other, loosely enough so as to allow free rotation).
\item Wiring.
\item Spektrum DX7S Radio.
\item Rod end, aluminium strips and pipe (for a stand, for testing).
\end{enumerate}

\section{Design for the Control of a Rotatable Stabiliser}

This section sets out to design a protocol for the control of a rotatable
stabiliser, such that its in-flight orientation and the deflection of its control surface can be actively adjusted in an intuitively obvious manner.

\subsection{Desired Mapping and its Implied Control Protocol}

Since the configuration of a rotatable stabiliser involves two degrees of
freedom, namely its orientation and the deflection of a hinged control surface
(or its angle of attack, for an all-moving, stabilator-like device), any
two-axis device will suffice as a controller. For succinctness, a mapping
betweeen states of the stabiliser and the position of the simplest two-axis
controller (a joystick) is formulated. The function of the joystick
traditionally assigned to the control of the ailerons is replaced by that
traditionally associated with rudder pedals, for simplicity and ease of
understanding. Its pitch-controlling function is otherwise conventional. There
is no loss of generality in making this joystick-only simplification. Pedal
inputs can just as easily be substituted for the lateral movements of a
joystick. The mapping is readily converted to the more conventional controls for
pitch and yaw, in either manned aircraft, or radio-controlled aircraft. What
follows is a description of how a continuous, one-to-one, conformal mapping,
between the position of controls and the state of the rotatable stabiliser, can
be devised in order for it to be controlled logically, reflexively and in an
elementary and intuitively obvious manner. 

\subsubsection{Embedded Substructures Within the Domains}

Convention dictates that the joystick retains the traditional pitch-altering
function of the aircraft. This requires that in the absence of any lateral
displacement of the joystick, the orientation of the stabiliser and the
deflection of its control surface should correspond to those of a horizontal
stabiliser and its elevator, respectively. The preferred orientation of the
rotatable stabiliser is therefore parallel to the lateral axis of the aircraft
for pitch-only inputs from the joystick. What is traditionally the banking
function of the joystick is, however, replaced by the yaw-adjusting function
traditionally assigned to the rudder pedals (this is the traditional mode four
on most radio-control apparatus). This means that for purely lateral
displacements of the joystick, the orientation of the stabiliser and the
deflection of its control surface correspond to those of a vertical stabiliser
and rudder, respectively. 

\begin{figure}[H]
    \begin{center}
\includegraphics[width=15cm, angle=0, clip = true]{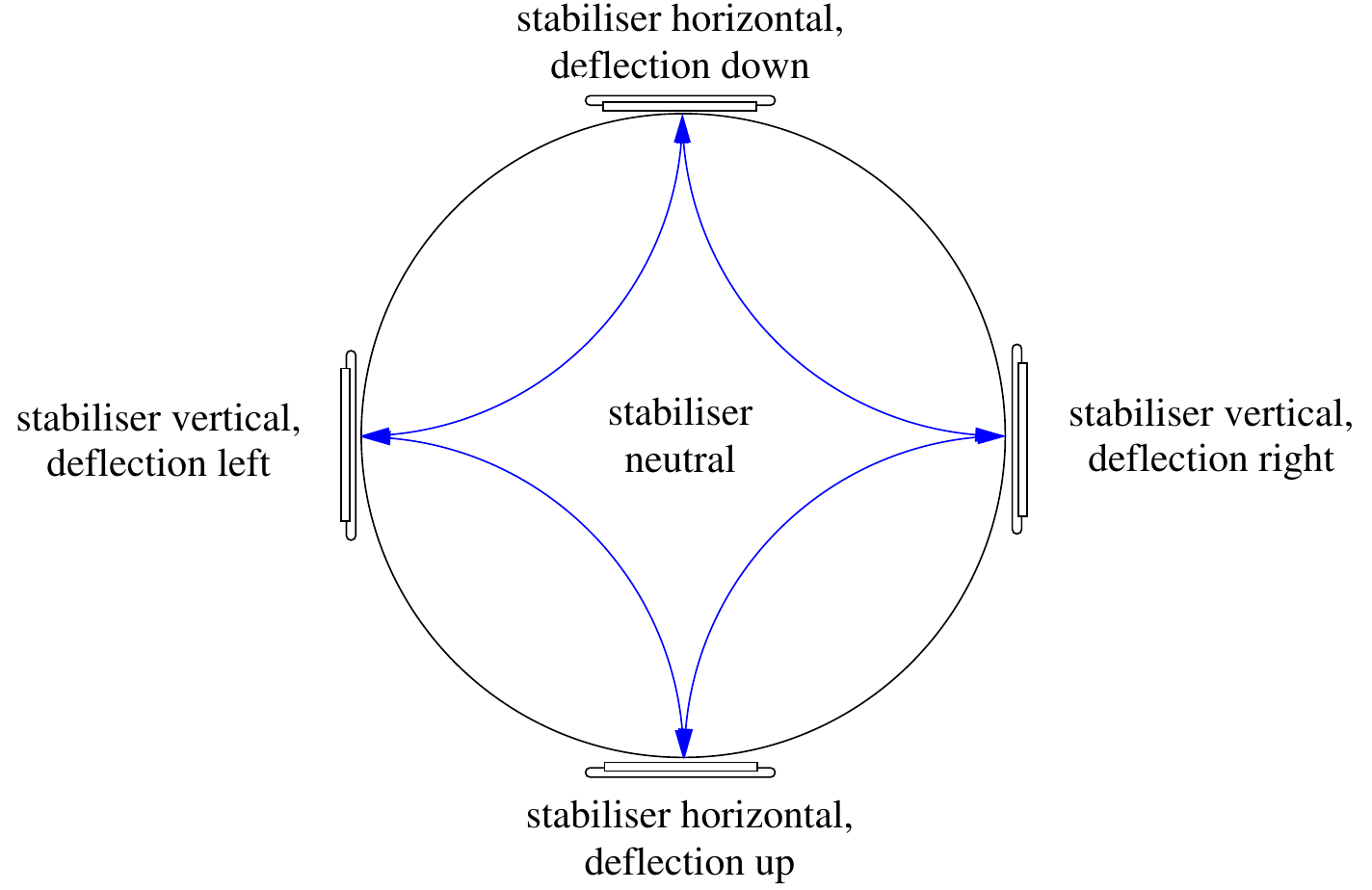}
\caption{The {\bf aft, or tail implementation} of a rotatable stabiliser. The
circle represents the boundary of the domain of any two-axis controlling device
e.g. a joystick, or a combination of one axis of a joystick and pedals.
Superimposed at four positions on it are the corresponding states of a rotatable
stabiliser as seen from aft. Blue arrows indicate the sense of rotation (no
translation) between the respective states.} \label{joystick}
   \end{center}
\end{figure}

Fig. \ref{joystick} relates the orientation of a rotatable stabiliser and the
deflection of its hinged control surface to positions of the controls. The
circle in Fig. \ref{joystick} represents the boundary of the domain of any
two-axis controlling device e.g. a joystick, or a combination of one axis of a
joystick and pedals. Superimposed at four positions on it are the corresponding
states of a rotatable stabiliser as seen from aft. Starting from
the top of Fig. \ref{joystick} and moving clockwise: Stick forward, nose pitches
down; right pedal or stick, nose yaws right; stick back, nose pitches up; left
pedal or stick, nose yaws left. 

\subsubsection{The Control Protocol which Stems from the Embedded Substructures} \label{rules}

A logical outgrowth of the two embedded substructures (Fig. \ref{joystick}) is
that small and continuous adjustments cause the stabiliser to rotate in a
direction opposite to that of the joystick (when viewed from aft) and the
deflection of the hinged control surface is proportional to the radial
displacement of the joystick from its centred position. The blue arrows in Fig.
\ref{joystick} indicate the sense of rotation (no translation) between the
respective states. 

For movements of the controls through the origin, or close to it, it may be
preferable to reverse the sense of deflection, rather than have the device
rotate excessively. Unproductive and excessive rotation due to small, repetitive
corrections and over-corrections can be avoided, in this way. Unproductive and
excessive rotation can readily be defined as adjustments of the joystick which
require rotation greater than 90~$^\circ$. The proposed remedy is to rotate the
device as if its orientation differs by 180~$^\circ$ and to reverse the
deflection to be the negative of what it otherwise would have been. Deflection
of the hinged control surface is, therefore, not always to the same side,
requiring the use of a symmetrical aerofoil.

A more exact formulation of the control protocol follows. Suppose $x$ and $y$
are used to denote the horizontal displacement of a joystick and, $\phi$, its
angular coordinate. Notice that, if $\Delta \phi$ is used to denote the change
in this angular coordinate ($\Delta \phi = \phi - \phi_{\mbox{\scriptsize
old}}$), then $\phi$ can always be converted to a periodic equivalent, by
implementing
\begin{eqnarray*}
\begin{array}{l} \phi -  2 \pi \\
\phi + 2 \pi \end{array} &\mbox{while}& \begin{array}{l} \Delta \phi > \pi \\ \Delta \phi < - \pi, \end{array}
\end{eqnarray*}
recursively,  to insure that \mbox{$\left|\left| \Delta \phi \right|\right| \le \pi$}. The proposed deployment of the device can then be formulated as follows:
\begin{enumerate}
\item The orientation of the aerofoil and the deflection of its control surface correspond to those of a horizontal stabiliser and elevator, respectively, for positions of the joystick on the line $x=0$. The orientation of the stabiliser and the deflection of its control surface correspond to those of a vertical stabiliser and rudder, respectively, for positions of the joystick on the line $y=0$ and $x\neq0$. 
\item For small and continuous adjustments of the joystick (the category
$\left|\left| \Delta \phi \right|\right| \le \frac{\pi}{2}$) the aerofoil
rotates in the opposite direction to the joystick, when the former is viewed
from aft. The relation
\begin{eqnarray*}
\theta &=& \frac{\pi}{2} - \phi \hspace{10mm} \mbox{for} \hspace{10mm} \phi \in \left[ \left. - \frac{\pi}{2}, \frac{3 \pi}{2} \right) \right.
\end{eqnarray*}
is one in which the orientation, $\theta$, of the aerofoil's lift force can be rotated evenly to starboard and port. 
\item The deflection of the hinged control surface (or the angle of attack in the case of a stabilator-like implementation) is proportional to
the radial displacement of the stick from the centred position. If $d$ is used to denote the deflection of the control surface, then it is related to the position of the joystick by
\begin{eqnarray*}
d &=& \frac{R}{r} \sqrt{x^2 + y^2},
\end{eqnarray*}
in which $\frac{R}{r}$ is some constant which callibrates the deflection of the aerofoil to the displacement of the joystick.
\item For adjustments of the joystick which militate a rotation greater than
90~$^\circ$ in terms of the aforementioned protocol, expedience dictates that it be applied as if the actual orientation of the device differs by
180~$^\circ$ from what it actually is and the sign of deflection be reversed. That is, if $\left|\left| \Delta \phi \right|\right| > \frac{\pi}{2}$, then
\begin{eqnarray*}
\theta &=& \left\{ \begin{array}{l} \frac{\pi}{2} - \phi + \pi \\ \frac{\pi}{2} - \phi - \pi \end{array} \right. \ \mbox{for} \ \begin{array}{l} \Delta \phi > 0 \\ \Delta \phi < 0 \end{array} \hspace{10mm} \mbox{and} \hspace{10mm} d \ = \ - \frac{R}{r} \sqrt{x^2 + y^2}.
\end{eqnarray*}
The above strategy becomes clearer when contemplating the alternative formulation, \begin{eqnarray*}
\theta &=& \theta_{\mbox{\scriptsize old}} - \Delta \phi \pm \pi,
\end{eqnarray*} 
in which $\theta_{\mbox{\scriptsize old}}$ is the old orientation of the aerofoil.
\end{enumerate}

Notice that however tempting it may be to exploit $\Delta \phi$ as a variable
with which to further simplify these formulae, the idea is to rapidly
recalculate the stabiliser's configuration (at 25 times per second), before the
previously calculated configuration has actually been achieved. The above
formulae were incorporated into a programme that was written using the Microchip
MPLAB IDE software, which utilizes the Microchip XC8 C compiler. The
microcontroller was programmed and debugged using the Microchip ICD2 debugger.
The debugger enabled the programme to be stepped through line by line.

In the conventional tail, two linearly independent inputs are mapped to two
linearly independent outputs, which combine to give a resultant. In the
rotatable-stabiliser concept the two linearly independent inputs are mapped
directly to a resultant. Notice that the transformation between states is
continuous, conflicting, successive control inputs being the exception. Clearly,
there would be a fundamental loss of continuity in the aforementioned diagrams,
were a tilting, as opposed to a fully rotating device to be used. 

Mistakenly rotating such a device into the vertical position represents a very
real danger at low altitudes. One, contemplated safety precaution was to
restrict the $x$-axis input so that the device is never tilted at an angle
greater than e.g. 27~$^\circ$ to the horizontal. This can easily be accomplished
by implementing the formula 
\begin{eqnarray*}
- \left|\left| \frac{y}{{\mathop {\rm tan}} \ 63 \ ^\circ} \right|\right| \le x \le \left|\left| \frac{y}{{\mathop {\rm tan}} \ 63 \ ^\circ} \right|\right|, 
\end{eqnarray*}
to restrict and modify the $x$-input from the joystick.

Pedal inputs can just as easily be substituted for the lateral movements of a
joystick, thereby converting the mapping to one associated with the more
conventional controls for pitch and yaw in manned aircraft. Likewise, the
mapping can also be readily converted to one associated with the more
conventional controls for pitch and yaw in radio-controlled aircraft, simply by
switching the transmitter, from its mode 4, to its mode 2. 


\section{Rapid Rotation at Low Speeds and its Implications for the Angle of Attack}

One, anticipated handicap of a rotatable stabiliser is the potential for it to
stall, from its tips, inward, if rotated too fast. Once rotation is underway,
the angle of attack is no longer simply that between the chord and the remote,
free air flow. Under such circumstances the incident air acquires an additional
component of velocity, opposite to the direction of rotation. Rapid rotation of
the device could therefore be a complication in slow moving aircraft. Rotation
of the device at maximum deflection (the stalling angle) could, likewise, be
predicted to be problematic. The outer tip of the aerofoil will begin to stall
during rotation, should rotation cause the maximum angle of attack to be
exceeded. Either the aerofoil must be rotated at a slower speed or a
differential angle of attack must be added and subtracted from either side of
the axis of rotation. 

The deviation in the angle of attack, $\Delta \alpha$, brought about by rotation is readily calculated according to the formula
\begin{eqnarray} \label{changeInAngleOfAttack}
\Delta \alpha &=& {\mathop {\rm arctan}}\left(\frac{\mbox{tangential velocity}}{\mbox{airspeed}}\right) \nonumber \\
&=& {\mathop {\rm arctan}}\left(\frac{\mbox{revolutions per second} \times 2 \pi \times \mbox{radius in $\mathrm{m}$}}{\mbox{airspeed in~$\mathrm{km} \ \mathrm{h}^{-1}$} \times \frac{10}{36}}\right), 
\end{eqnarray}
in which the radius referred to is the distance along the aerofoil from the
point of rotation. From this formula one immediately observes that short spans
are conducive of small deviations in the angle of attack, as are the kind of
high airspeeds one normally associates with missiles and their like. The area,
hence lift, lost in shortening the span can, to a certain extent, be recouped by
means of a longer chord. 

For a radio controlled aircraft with a rotatable stabiliser of 40~$\mathrm{cm}$
and a 16~$\mathrm{km} \ \mathrm{h}^{-1}$ stall speed, one would expect rotation
at a rate of $\frac{\pi}{2}$~$\mathrm{s}^{-1}$ to induce a departure from the
angle of attack, at the tips, which would never exceed $\pm$ 4~$^\circ$. What
kind of stalling angles are contemplated? Symmetrical aerofoils with a high
stalling angle, such as the NACA 0015, are a common choice for stabilators. In
theory this aerofoil stalls just above 22~$^\circ$ while simultaneously
delivering a lift coefficient just greater than 1.5 and a lift-to-drag ratio
slightly above 95 (Jacobs et al. \cite{jacobsWardPinkerton:1}). In the
real world induced drag and atmospheric conditions, e.g. wind shear and gusting,
can dramatically reduce this angle and the margin of safety. The functional
range of angles of attack for aerofoils, in general, is usually cited as being
in the -4--16~$^\circ$ range by more practical references concerned with less
ideal conditions (e.g. Thom \cite{trevorThom:1}). It was decided to set
the rate of aerofoil rotation to just over  $\frac{\pi}{2}$~$\mathrm{s}^{-1}$
(15 $\mathrm{rpm}$) in the radio-controlled, model aircraft.

\section{Testing the Rotatable Stabiliser}

\begin{figure}[H]
    \begin{center}
        \includegraphics[width=15cm, angle=0, clip = true]{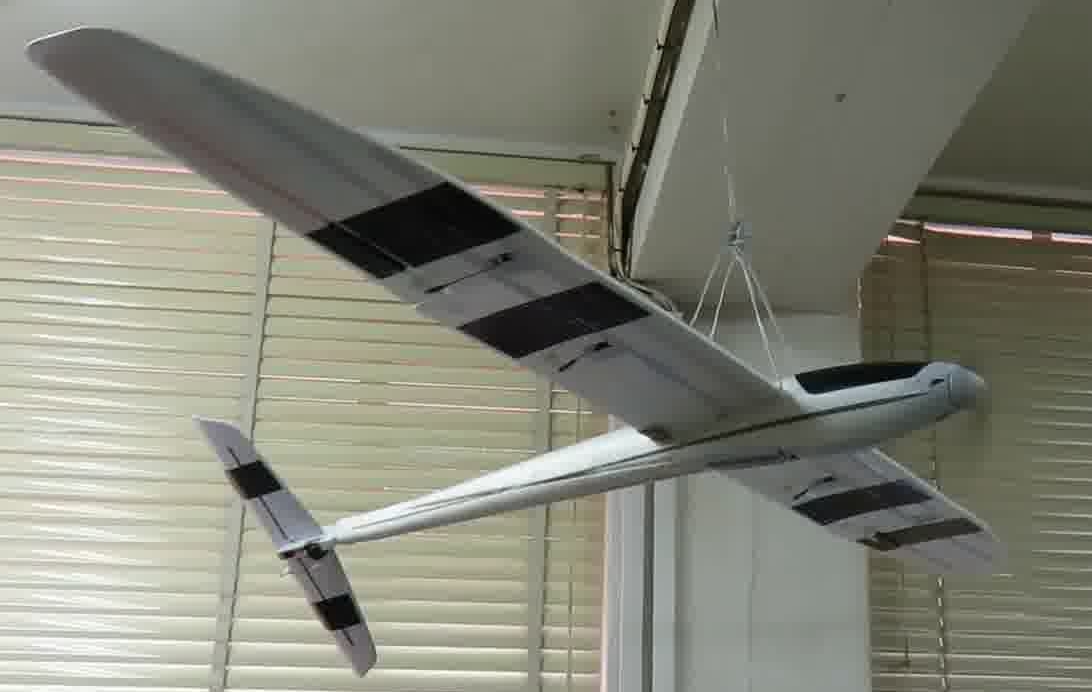}
	\caption{The completed modification of the radio-controlled, model aircraft initially being tested in its own propwash, while suspended from a tether.}
\label{completedAeroplane}
   \end{center}
\end{figure}

The first stage of testing involved the completed aircraft simply being
suspended from a tether and being subjected to its own propwash (refer to Fig.
\ref{completedAeroplane}). 
Thereafter the model aircraft, sans wings, was attached by way of a cradle, to a
rod-end (or universal joint), thereby allowing it to be either mounted above a
car (driven at around 30~$\mathrm{km} \ \mathrm{h}^{-1}$) or placed in a
make-shift wind tunnel (the entrance to a hospital air-conditioning unit). This
was to test and familiarise the authors with the controls. The third stage of
tests involved re-attaching the wings and flying the aircraft. The device was
initially tested using mode 4 of the radio. This was subsequently changed to
mode 2 for the flight tests.

\section{Results}

The modified, radio-controlled aircraft flew successfully and its performance
lived up to the authors' best expectations. Surprisingly, no side effects
arising from the omission of a vertical stabiliser were evident and the flight
of the modified, model aircraft did not seem to be impaired by its absence. The
drag on the tail seemed to be sufficient for the purposes of static longitudinal
and directional stability.

\begin{figure}[H]
    \begin{center}
        \includegraphics[width=14.5cm, angle=0, clip = true]{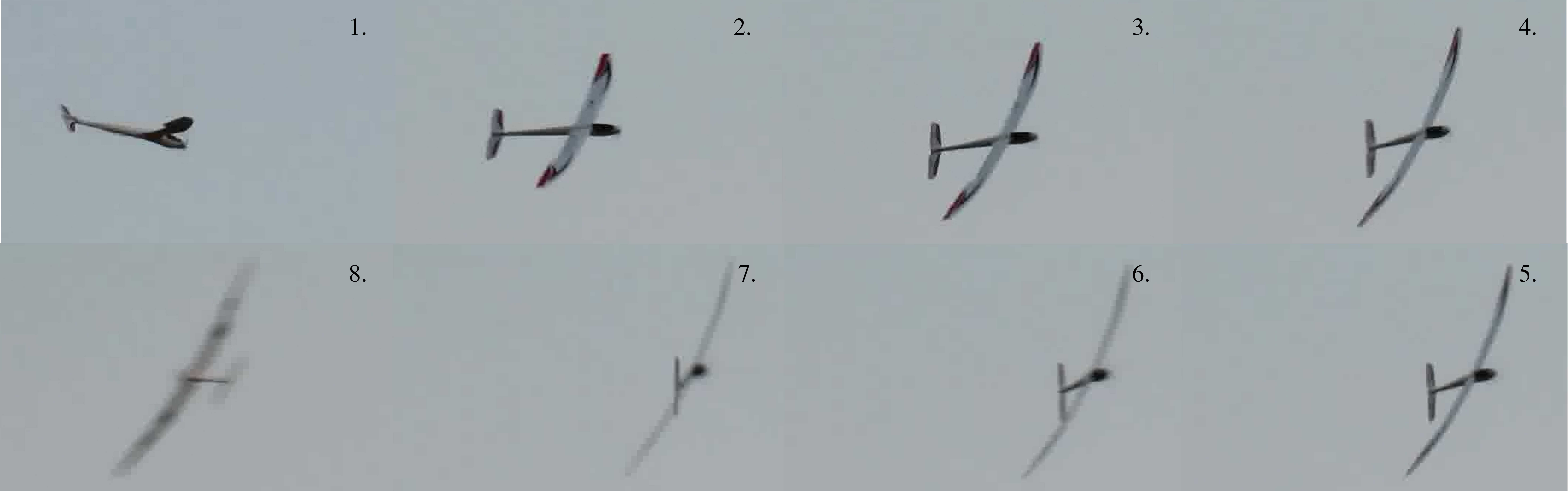}
	\caption{A sequence showing the modified, radio-controlled aircraft turning to starboard.}
\label{sequence13}
   \end{center}
\end{figure}

\begin{figure}[H]
    \begin{center}
        \includegraphics[width=14.5cm, angle=0, clip = true]{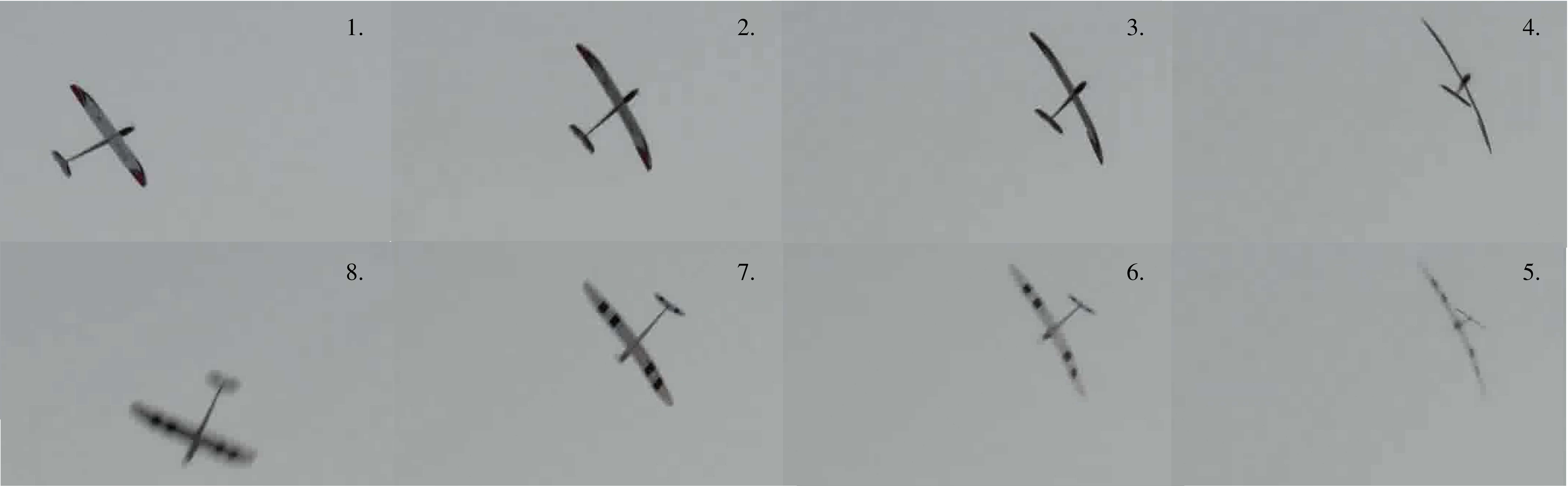}
	\caption{A sequence showing the radio-controlled aircraft making a very steep, aerobatic turn to starboard.}
\label{sequence12}
   \end{center}
\end{figure}

\begin{figure}[H]
    \begin{center}
        \includegraphics[width=14.5cm, angle=0, clip = true]{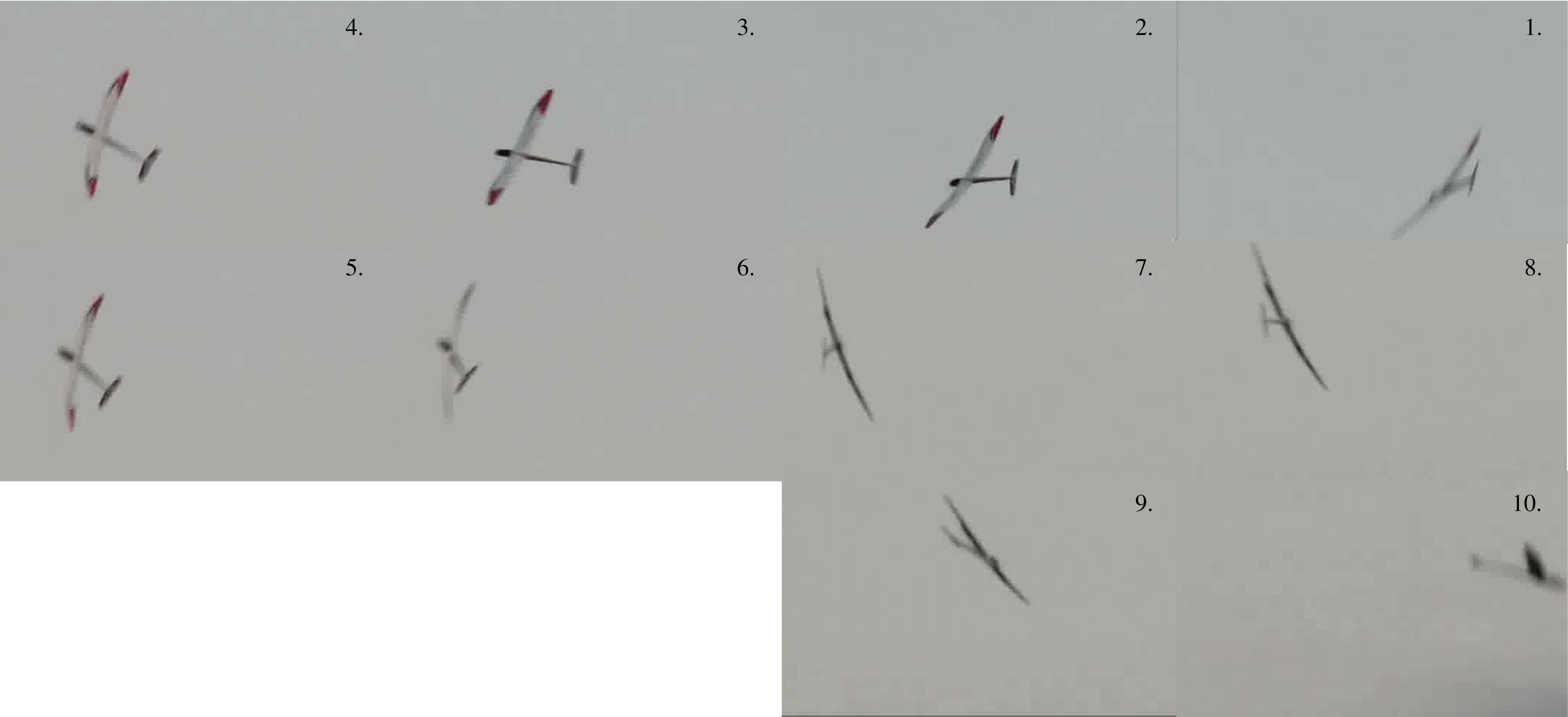}
	\caption{A sequence showing the modified, radio-controlled aircraft turning to port.}
\label{sequence1}
   \end{center}
\end{figure}

\begin{figure}[H]
    \begin{center}
        \includegraphics[width=14.5cm, angle=0, clip = true]{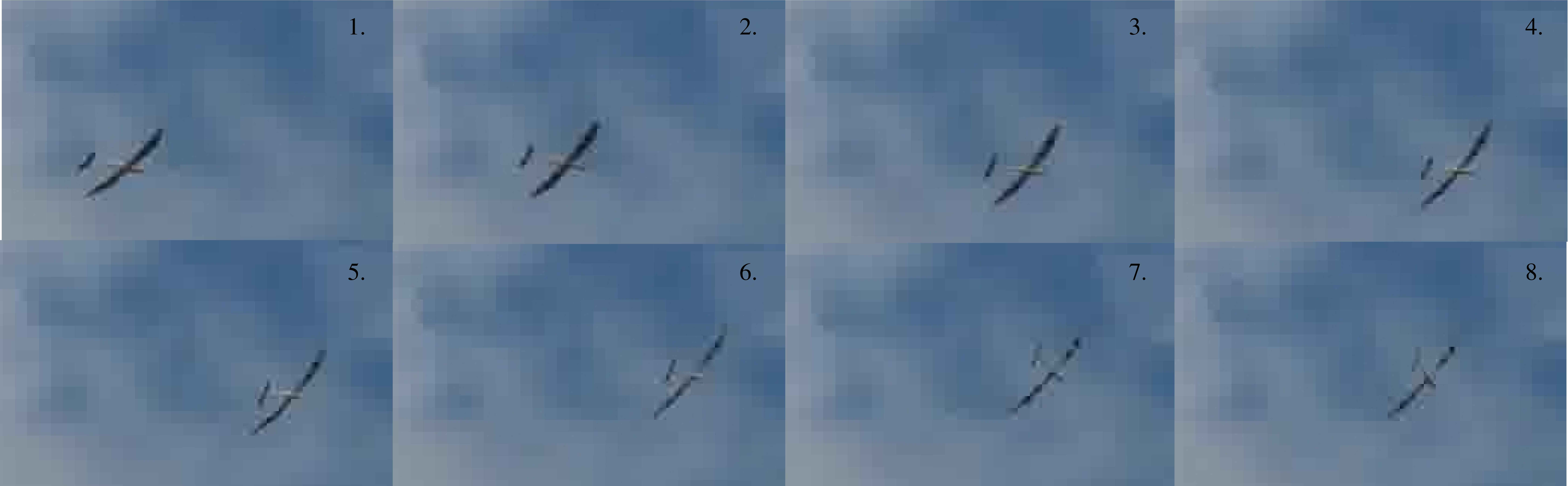}
	\caption{A sequence showing a purely yaw-induced bank to port (no aileron).}
\label{sequence15}
   \end{center}
\end{figure}
The orientation of the stabiliser is clearly visible, as the aircraft turns to
starboard, in the \mbox{Fig. \ref{sequence13}}, clockwise sequence. The
orientation of the stabiliser is also clearly visible, as the the aircraft
performs a steeper, aerobatic turn to starboard, in the Fig. \ref{sequence12},
clockwise sequence. The \mbox{Fig. \ref{sequence1}}, winding sequence depicts a
turn to port. The Fig. \ref{sequence15} sequence shows a purely yaw-induced bank
to port (no aileron), at the end of which the control surface unexpectedly
detaches under the stress.

%
\section{Conclusions}

This research implements a design for the control of a rotatable stabiliser
which, it is proposed, might augment, or fully replace, the conventional
mechanisms for pitch and yaw in certain types of aircraft. The anticipated
advantages of such a device are around 25\% less drag, for a capability which
ranges between equivalent, to greater than twofold that of the conventional tail
(refer to Eq. \ref{equivalentArea} and Fig. \ref{envelopeMaxForce}).
Interference drag should also be eliminated.

Whether the controls are a single joystick, two joysticks, or a combination of
joystick and pedals as in manned aircraft, their relevant position may be
described by two coordinates, $x$ and $y$. A change in their angular coordinate
can always be expressed in a format $\left|\left| \Delta \phi \right|\right| \le
\pi$, by implementing
\begin{eqnarray*}
\begin{array}{l} \phi -  2 \pi \\
\phi + 2 \pi \end{array} &\mbox{while}& \begin{array}{l} \Delta \phi > \pi \\ \Delta \phi < - \pi, \end{array}
\end{eqnarray*}
recursively, for $\phi$ initially assumed to be from $\left[ \left. - \frac{\pi}{2}, \frac{3 \pi}{2} \right) \right.$. The two modes of control implemented are then as follows: If $\left|\left| \Delta \phi \right|\right| \le \frac{\pi}{2}$, the orientation of the aerofoil, $\theta$, and the deflection, $d$, of its control surface are calculated according to
\begin{eqnarray*}
\theta &=& \frac{\pi}{2} - \phi \hspace{10mm} \mbox{and} \hspace{10mm} d \ = \ \frac{R}{r} \sqrt{x^2 + y^2}.
\end{eqnarray*}
Otherwise, if $\left|\left| \Delta \phi \right|\right| > \frac{\pi}{2}$, the configuration of the stabiliser is calculated according to
\begin{eqnarray*}
\theta &=& \left\{ \begin{array}{l} \frac{\pi}{2} - \phi + \pi \\ \frac{\pi}{2} - \phi - \pi \end{array} \right. \ \mbox{for} \ \begin{array}{l} \Delta \phi > 0 \\ \Delta \phi < 0 \end{array} \hspace{10mm} \mbox{and} \hspace{10mm} d \ = \ - \frac{R}{r} \sqrt{x^2 + y^2}.
\end{eqnarray*}
The result is that, for a pitch-only input, the orientation of the stabiliser
and the deflection of its control surface correspond to those of a horizontal
stabiliser and its elevator, respectively. Conversely, for a yaw-only input, the
orientation of the stabiliser and the deflection of its control surface
correspond to those of a vertical stabiliser and rudder, respectively. Small and
continuous adjustments cause the aerofoil to rotate in the opposite direction to
the controls (when viewed from aft) and the deflection of the hinged control
surface is proportional to the radial displacement of the joystick from its
centred position. For what, in terms of that protocol, would amount to large and
contradictory rotations of the aerofoil (rotations greater than 90~$^\circ$) the
desired configuration is more efficiently achieved by regarding its original
orientation to differ by 180~$^\circ$ from what it actually is and by reversing
the sign of the deflection. This second mode of control obviously implies the
use of a symmetrical aerofoil, such as the NACA 0015. This should not be a cause
for concern, however, since the NACA 0015 has an almost identical performance to
that of the NACA 2412 (Jacobs et al. \cite{jacobsWardPinkerton:1}), the
latter being the preferred choice in many Cessnas. The algorithm to determine
the aerofoil's orientation and the deflection of its control surface was run 25
times per second.

One, anticipated handicap of such a device is the potential for it to stall,
from its tips, inward, if rotated too rapidly. The deviation in the angle of
attack, $\Delta \alpha$, brought about by a rapid adjustment can be calculated
according to the formula
\begin{eqnarray*}
\Delta \alpha &=& {\mathop {\rm arctan}}\left(\frac{\mbox{revolutions per second} \times 2 \pi \times \mbox{radius in $\mathrm{m}$}}{\mbox{airspeed in~$\mathrm{km} \ \mathrm{h}^{-1}$} \times \frac{10}{36}}\right), 
\end{eqnarray*}
in which the radius referred to is the distance along the aerofoil from its axis
of rotation. A restriction on the speed of rotation therefore exists and a
strategy in which the device is never rotated with its control surface at
maximum deflection is obviously prudent. The rate of the aerofoil's rotation was
set to just over $\frac{\pi}{2}$~$\mathrm{s}^{-1}$ (15 $\mathrm{rpm}$) in the
radio-controlled model. This was slow enough to prevent any noticeable, outboard
stalling, yet fast enough to prevent any noticeable lag in control. No problems
were encountered in the radio-controlled model, however, the danger of a
large-span stabiliser stalling, from its tips, inward, at low airspeeds during
rotation, make the device ideally suited to high speed aircraft, missiles and
drones. The former pair are both craft in which either a short tail-span or fins
are preferred. A high-performance response is usually not required from drones
and their tendency not to engage in rapid aerobatic manoeuvres should permit
slow rotation, making them just as likely candidates for the implementation of a
rotatable stabiliser. In a manned aircraft the response of the normal tail
configuration is instantaneous, at least in so far as the human input is. In a
remotely controlled aircraft one relies on servos, anyway. Setting a different
pitch on either side of the axis of rotation, or even incrementally warping the
aerofoil along its full length, should that be possible, would be one counter
measure to the problem of stalling at the tips during rotation. 

No weathervaning of the tail, something which ultimately manifests itself as
Dutch roll, was observed in the many flights of the radio-controlled, model
aircraft. It should, however, be emphasised that the model aircraft's wing had
no, or little sweepback (curved along the leading edge) and a small amount of
in-flight dihedral. The radio-controlled aircraft's small scale, low speed and,
consequently, the low Reynolds number of the airflow around it severely limit
the extrapolation of these results to large, fast aircraft, missiles and even
drones. Although the radio-controlled aircraft appeared to demonstrate good,
inherent, static longitudinal and directional stability, intensive
experimentation would obviously be required before the deployment of such a
device in any civil aircraft could be contemplated. 

The ever-present danger of a pilot inadvertantly making a yaw-only input is a
real concern. Rotating the aerofoil into the vertical at low level has obvious
consequences. Devising and programming a precautionary control strategy against
such an eventuality is, nonetheless, no insurmountable obstacle, to
the extent it might, more aptly, be considered a topic for further
experimentation and debate, rather than one of research. A very minor,
nonetheless, important problem, a solution is necessary. The precautionary mode,
whereby the $x$-axis input is restricted to
\begin{eqnarray*}
- \left|\left| \frac{y}{{\mathop {\rm tan}} \ 63 \ ^\circ} \right|\right| \le x \le \left|\left| \frac{y}{{\mathop {\rm tan}} \ 63 \ ^\circ} \right|\right| 
\end{eqnarray*}
(so that the device is never tilted at an angle greater than 27~$^\circ$ from
the horizontal), had the unpleasant side-effect of disabling all yaw control,
from time to time. It was found to be too restrictive and abandoned. Another,
untested alternative would be to stipulate a minimum $y$-axis input, to be
triggered by yaw-only inputs. A myriad of options, with variations in a
conflated pitch input whose decay ranges from linear to exponential, exist. The
inherent danger obviously diminishes as an aircraft's design approaches that of
a so-called flying-wing (e.g. the radio-controlled Wombat, the more extreme
\mbox{Horten Ho 229} or \mbox{Northrop-Grumman X-47B}), envisaged by many to be
the future of aviation. In view of the alarm expressed at this work from certain
quarters, it should also be pointed out that the \mbox{Horten Ho 229} reportedly
only suffered from moderate lateral instability and videos of the
\mbox{Northrop-Grumman X-47B} are not suggestive of any instability at all. This
in the complete absence of any stabilisers at all! Wingtip modifications
designed to counter induced drag might also be predicted to diminish the
vertical stabiliser's significance in future aircraft designs. 

Although the modified servo's limitation of $\theta \le \left|\left| \pi
\right|\right|$ was never encountered during any of the test flights, that
limitation would easily be eliminated by the deployment of a stepper motor, in
conjunction with a shaft encoder, in a slightly larger aircraft. A limit of
finite rotation could be a very real risk to any aircraft of value. Were a
tilting, as opposed to a rotating device to be used, there would clearly be a
fundamental loss of continuity in the movement of the device and therefore the
control of the aircraft. In the case of a 90~$^\circ$-tilting device, for
example, one cannot effect a continuous transition between a nose-up to
nose-down input, while simultaneously maintaining a yaw input, as this results
in a lateral inversion of the yaw. 

Although it is contended that the mapping betweeen states of the device and the
position of a single, two-axis controller (e.g. a joystick) is the most logical,
it was found that even the best radio-pilots could not overcome their reflexes.
The mapping was readily converted to the more conventional controls for pitch
and yaw in radio-controlled aircraft; that is, implemented under the standard
mode two of the radio, rather than its mode four. Pedal inputs can similarly be
substituted for the lateral movements of a joystick. 

The Parkzone Radian Pro was selected on the basis of it being fairly ubiquitous.
This may not have been the most astute choice, in retrospect, since its
flexibility often led to an elastic response and the tailboom meant that the
stabiliser had to be placed close to downwash. It would also have been
preferable not to have positioned the device in the propwash. The Skysurfer may
have been a better choice.

The only difference between a canard-wing type deployment and the tail
implementation which is the topic of this work, would be that the deflection of
the hinged control surface is in the opposite direction. For a stabilator-like
device, the specified deflection relates to the trailing edge, instead of a
hinged control surface. In helicopters the tail consists of a rotor with pitch
and speed control, which complicates matters only very slightly. In such
circumstances the direction of the deflection can be taken to specify the
direction of thrust, instead. While the concept of directed lift is presently
developed in the context of an aerofoil, an analogous description for a rotor is
easily deduced. If one were to coin a term for a whole family of such devices
e.g. `swivelator' it would refer to an aerofoil (or rotor) whose in-flight
orientation is actively adjusted by
swivelling\footnotemark[3]\footnotetext[3]{Note the capability implied by the
use of the operative word ``swivelling'' as opposed to ``tilting''.}, relative
to the aircraft, around an axis parallel, or approximately parallel, to the
longitudinal axis, thereby affording the device the capability of exerting a
lift force in any direction about that axis. The magnitude of the lift force is
designed to be adjustable by way of a hinged control surface, changing the angle
of attack (a l\'{a} a stabilator), or additionally, in the case of a rotor
blade, spinning faster or changing the pitch of the rotor blades. Aileron-like
inputs might also be superimposed (as in elevons) in implementations where the
rotatable stabiliser has separate control surfaces on either side of the axis
about which it rotates.


\section{Acknowledgements} 

Michael Mettler, Pieter Botes and Adriaan Hugo provided work-shop assistance.
The authors are indebted to Reynard Myburgh and Danie Krugel for assistance in
putting the radio-controlled, model aircraft through its paces. Johan Meyer and
Glen Taylor are also deserving of thanks. 

%
%
\bibliography{rotatableStabilator}

\begin{thebibliography}{1}

\bibitem{patentChilds11}
S.~Childs.
\newblock {\em Swivelator}.
\newblock Provisional patent 2004/7616. 2004.

\bibitem{patentChilds12}
S.~Childs.
\newblock {\em Swivelator}.
\newblock Provisional patent 2004/10179. 2004.

\bibitem{patentChilds13}
S.~Childs.
\newblock {\em Swivelator}.
\newblock Provisional patent 2005/05287. 2005.

\bibitem{jacobsWardPinkerton:1}
E.N. Jacobs, K.E. Ward, and R.M. Pinkerton.
\newblock The characteristics of 78 related airfoil sections from tests in the
  variable-density wind tunnel.
\newblock Technical Report 460, National Advisory Committee for Aeronautics,
  1933.

\bibitem{trevorThom:1}
T.~Thom.
\newblock {\em The Aeroplane -- Technical}, volume~4 of {\em The Air Pilot's
  Manual}.
\newblock 1993.

\end{thebibliography}

\section*{Appendix}

\subsection*{Notation}

\begin{table}[H]
\begin{center}
\begin{tabular}{c l}  
& \\
Symbol \ & \hspace{25mm} Description \\ 
 \ & \\
$C_L$ \ & \ coeficient of lift \ \\
$d$ \ & \ deflection of the hinged control surface \ \\
$\frac{R}{r}$ \ & \ constant callibrating the deflection of the aerofoil \ \\ 
\ & \ to the displacement of the joystick \ \\
$S$ \ & \ area \ \\
$v$ \ & \ velocity of the incident air \ \\
$x$ \ & \ lateral displacement of the joystick, or pedal input \ \\
$y$ \ & \ longitudinal displacement of the joystick \ \\
$\Delta \alpha$ \ & \ change in the angle of attack \ \\
$\theta$ \ & \ angle between the stabiliser and the lateral axis \ \\
$\theta_{\mbox{\scriptsize old}}$ \ & \ old orientation of the stabiliser \ \\ 
$\rho$ \ & \ density of the air \ \\
$\phi$ \ & \ angular coordinate of the joystick's position \ \\
$\phi_{\mbox{\scriptsize old}}$ \ & \ previously recorded angular coordinate of the joystick \ \\
\end{tabular}
\end{center}
\end{table}

\subsection*{Algorithm}

The programme for the control of the rotatable stabiliser, written in C, is as follows.

{\tiny

\noindent 

\begin{lstlisting}
//******************************************//

// Servo controller
// Rotatable Stabiliser
// V0.8 - changed to Simon's code
// 26 August 2015
// Mark Jackson
//
//******************************************//

#include <xc.h>								//Include file for the compiler
#include <math.h>							//Include file for math functions

//******************************************//
__CONFIG(									//Configuration bits for processor
		FOSC_HS &							//High speed oscilator
		WDTE_OFF & 							//Watchdog off
		PWRTE_OFF & 							//Power-up timer off
		CP_OFF & 							//Code protect off
		LVP_OFF & 							//Low voltage programing ogg
		DEBUG_OFF & 						//Debug mode off
		CPD_OFF & 							//Code protect data off
		BOREN_OFF);							//Brownoutreset off
											
//******************************************//

#define _XTAL_FREQ 		16000000			// Define clock speeed.
#define rud_in  	    RB5					// Rudder signal input.
#define elev_in 	    RB4					// Elevator signal input.
#define rud_out  	    RB1					// Orientation signal output.
#define elev_out     	RB0					// Deflection signal output.

//******************************************//

float 			rudder;   					// `Rudder', or x-input.
float			elevator;					// `Elevator', or y-input.
int 			rud_new;					// New rudder value.
int 			elev_new;					// New elevator value.
int 			rud_newnew;					// New rudder value.
int 			elev_newnew;					// New elevator value.

float 			rud_temp;					// New rudder value.
float 			elev_temp;					// New elevator value.
int				offset;						// Offset value to zero captured signal values
float			pi = 3.14159265359;			// The value of pi.
float 			inv_pi = 0.31830988618;		// The reciprocal of pi.
signed int		deflection = 0;	 			// The value of the deflection output.
signed int		deflectionOld = 0;	 			// Value for deflection output
signed int 		deflectionCoefficient = 1;	// The coefficient for deflection reversal.	
int				tolerance = 4;				// Value for `dead' zone surrounding the origin.
float 			phi;						// Angular coordinate of the joystick.
float			deltaPhi;					// Change in the angular coordinate of the joystick.
float			phiOld = 90;				// Previous angular coordinate of the joystick.
float			theta = 0;					// Aerofoil orientation.
float			thetaPre = 0;					//
float			thetaOld;					// Previous aerofoil orientation.
float			mode2Addition = 0;          // The accumulated re-orientations of the phi reference.
signed int		orientation = 0;
char 			t;							//Temp value used to clear RBIF
char			rot_speed = 10;				//value that determines rotation speed
signed char		def_val = 0;				//value for deflection offset
char			startup;
char			ot = 2;
char 			centre = 0;
float			tan63 = 1.9626;
signed float	rud_tan63;					// Tan of 63 degrees.

//******************************************//

void main()									//
{											//configure proccesor
	PORTA = 0x00;							//
	PORTB = 0x00;							//
	TRISA = 0xFF;							//
	TRISB = 0x30;							//
	OPTION_REG = 0x88;						//
	INTCON = 0x48;							//
	PIE1 = 0x00;							//
	PIE2 = 0x00;							//
	T1CON = 0x20;							//
	ADCON0 = 0x00;							//
	ADCON1 = 0x87;							//
	PORTA = 0x00;							//
	PORTB = 0x00;							//
					
	TMR1H = 0x00;							//
	TMR1L = 0x00;							//
	INTF = TMR1IF = 0;						//
	RBIF = 0;								//
	GIE = 0;								//	

//******************************************//
//get centre offset
	startup = 0;
	while(startup == 0)
	{
	while(rud_in)							//wait for input to return to zero (logic 0)
	{}										//
	t = PORTB;								//read portb
	RBIF = 0;								//clear interrupt flag for change on input port
	while(RBIF == 0)						//wait until there is a change on a input pin
	{}										//
	if(elev_in)								//if elevator input is high (logic 1)
	{										//
	elev_newnew = 0;							//clear register
	while(elev_in)							//while elev input is high
	{										//
	elev_newnew++;								//increment register by one
	__delay_us(5);							//wait
	}										//
	t = PORTB;								//read portb
	RBIF = 0;								//clear interrupt flag
	}										//
	__delay_us(400);						///300
	while(RBIF == 0)						//same sequence as above for rudder input
	{}										//
	if(rud_in)								//
	{										//
	rud_newnew = 0;							//
	while(rud_in)							//
	{										//	
	rud_newnew++;								//	
	__delay_us(5);							//	
	}										//
	t = PORTB;								//
	RBIF = 0;								//
	}

	if(rud_newnew == elev_newnew)
	{
	offset = elev_newnew;
	startup = 1;
	}
	}

//******************************************//
//Main loop
while(1)
{
//*********//
//make outout signal for angle
	TMR1IF = 0;
	thetaPre = theta;
	orientation = (int)((-theta * 2.) + 1498);
//	orientation_temp = (orientation - orientation_old);		//used to slow down ratation speed
//	orientation_temp = (orientation_temp * orientation_temp);
//	orientation_temp = sqrt(orientation_temp);
//	if(orientation_temp > rot_speed)
//	{
//	if(orientation > orientation_old)
//	{orientation = orientation_old + rot_speed;}
//		if(orientation < orientation_old)
//	{orientation = orientation_old - rot_speed;}
//	}
//	orientation_old = orientation;
	orientation = (0xFFFF - orientation);
	TMR1L = orientation;
	orientation = orientation >> 8;
	TMR1H = orientation;
	TMR1ON = 0x01;
	rud_out = 1;
	while(TMR1IF == 0)
	{}
	rud_out = 0;
	theta = thetaPre;
	TMR1ON = 0x00;
	TMR1H = 0x00;
	TMR1L = 0x00;
//*********//
//make outout signal for deflection
	TMR1IF = 0;
	deflectionOld = deflection;
	deflection = (deflection * 14);
	deflection = (deflection + 1498);
	deflection = (0xFFFF - deflection);
	TMR1L = deflection;
	deflection = deflection >> 8;
	TMR1H = deflection;
	TMR1ON = 0x01;
	elev_out = 1;
	while(TMR1IF == 0)
	{}
	elev_out = 0;
	deflection = deflectionOld;
	TMR1ON = 0x00;
	TMR1H = 0x00;
	TMR1L = 0x00;
	RBIF = 0;
//*********//

//Capture pulses

	while(RBIF == 0)						// Wait until there is a change on an input pin.
	{}										//
	if(elev_in)								//if elevator input is high (logic 1)
	{										//
	elev_newnew = 0;							//clear register
	while(elev_in)							//while elev input is high
	{										//
	elev_newnew++;								//increment register by one
	__delay_us(5);							//wait
	}										//
	t = PORTB;								//read portb
	RBIF = 0;								//clear interrupt flag
	}										//

	while(RBIF == 0)						//same sequence as above for rudder input
	{}										//
	if(rud_in)								//
	{										//
	rud_newnew = 0;							//
	while(rud_in)							//
	{										//	
	rud_newnew++;								//	
	__delay_us(5);							//	
	}										//
	t = PORTB;								//
	RBIF = 0;								//
	}
	elev_new = elev_newnew;		
	rud_new = rud_newnew;

//*********//

	if(ot == 0)
	{

/* Restrict the rudder input as dictated by the precautionary mode, preventing the */
/* aerofoil tilting beyond 27 degrees. */

//	rud_temp = (int)( rud_new - offset );
//
//	elev_temp = (int)( elev_new - offset );	
//
//	elev_temp = ( sqrt( elev_temp * elev_temp ) / tan63 );
//	
//	if( rud_temp > elev_temp )
//		rud_tan63 = elev_temp;
//
//	else if( rud_temp < - elev_temp )
//		rud_tan63 = - elev_temp;
//
//	else
//		rud_tan63 = rud_temp;	
//
//	rudder = rud_tan63;
//
/* Calculate the delfection of the control surface. */

/* ... "offset" is the `origin' ... */

	rudder  = rud_new - offset;
	elevator = elev_new - offset;			//

	deflection = (int)sqrt( rudder * rudder + elevator * elevator ); 

	deflection = deflection + def_val;		//add offset value (new)

/* If the deflection is greater that xx, ... */

	if( deflection > 37 )						
		deflection = 37;

/* ... limit value to xx. */	


/* Calculate the orientation and mode of rotation (and deflection) of the aerofoil. */

/* If there is a meaningful input ... */

	if( deflection > tolerance )
	{
		thetaOld = theta;

		phiOld = phi;	

/* ... remember the previous orientation of the aerofoil and angular coordinate of
   the joystick. */
/* Determine the angular coordinate of the joystick ... */

		if( rudder > 0 )
			phi = atan( elevator / rudder );

		else if( rudder == 0 )
		{
			if( elevator > 0 )
				phi = pi / 2.;

			else
				phi = - pi / 2.;
		}

		else if( rudder < 0 )
			phi = pi + atan( elevator / rudder );

/* ... the line x = 0 and the second and third quadrants being the exception. */
/* Convert the angular coordinate of the joystick ... */

		phi = phi * 180 * inv_pi;	

/* ... to degrees. */
/* Choose the smallest of the two, possible angles through which to rotate, ... */
 
		while( ( phi - phiOld ) > 180 )
			phi = phi - 360;

		while( ( phi - phiOld ) < - 180 )
			phi = phi + 360;

/* ... exploiting the periodicity for the shortest possible route. */
/* Calculate the change in joystick angular coordinate, ... */ 

		deltaPhi = phi - phiOld;				

/* ... the criterion for choosing Mode 1 or Mode 2. */
/* If the first of the Mode 2 categories, ... */

		if( deltaPhi > 90 )
		{
			mode2Addition = mode2Addition + 180;

/* ... add 180 degrees to the accumulated, re-orientation term, ... */ 

			if( mode2Addition >= 360 )
				mode2Addition = mode2Addition - 360;

/* ... remove any periodicity, ... */ 

			deflectionCoefficient = - deflectionCoefficient;

/* ... and reverse the deflection. */ 
		}

/* If the second of the Mode 2 categories, ... */

		if( deltaPhi < - 90 )
		{
			mode2Addition = mode2Addition - 180;

			if( mode2Addition <= -360 )
				mode2Addition = mode2Addition + 360;

			deflectionCoefficient = - deflectionCoefficient;
		}

/* ... devise a analogous short-cut for the rotation. */
/* Finally, calculate the new orientation ... */

		theta = 90 - phi + mode2Addition;

		deflection = deflectionCoefficient * deflection;

/* ... and deflection of the aerofoil. */
	
		if(theta > 180)
			theta = 180;

		if(theta < -180)
			theta = -180;
	}
	else
	{
/* Otherwise, ... */

		centre++;

		if( centre > 2 )
		{
/* ... if there too long, ... */
			centre = 0;

			phi = 90;

			phiOld = 90;

			theta = 0;

			mode2Addition = 0;

			deflectionCoefficient = 1;
		}

/* ... set the centred values. */
	}

	ot = 4;
	}
	else
	{
	__delay_ms(4);	
	}
	ot--;
//*********//
}//while(1)
//******************************************//	
}//main
//******************************************//


 
\end{lstlisting}
}

\end{document}